\documentclass[aps,pre,twocolumn,groupedaddress,showkeys,showpacs]{revtex4}
\usepackage{amsmath}
\usepackage{amsfonts}
\usepackage{amssymb}
\usepackage{graphicx}

\newcommand{\mb}[1]{{\mathbf{#1}}}
\newcommand{\gb}[1]{\mbox{\boldmath{$#1$}}}
\newcommand{\prs}[1]{{\left(#1\right)}}
\newcommand{\chs}[1]{{\left\{#1\right\}}}
\newcommand{\col}[1]{{\left[#1\right]}}
\newcommand{\norm}[2]{{\mathcal{N}\prs{#1,#2}}}
\newcommand{\prob}[1]{{\mathcal{P}\prs{#1}}}
\newcommand{\sgn}{{\mbox{sgn}\,}}
\newcommand{\erf}{{\mbox{erf}\,}}
\newcommand{\erfc}{{\mbox{erfc}\,}}
\newcommand{\avg}[2]{{\left<#1\right>_{#2}}}
\newcommand{\iav}{{A,\mb{r},\mb{t}}}
\newcommand{\hpi}{{\hat{\pi}}}
\newcommand{\hx}{{\hat{x}}}
\newcommand{\bx}{{\mathbf{x}}}
\newcommand{\bhx}{{\mathbf{\hx}}}

\begin{document}

\begin{abstract}
The typical behaviour of the relay-without-delay channel and its
many-units generalisation, termed the \textit{relay array}, under
LDPC coding, is studied using methods of statistical mechanics. A
demodulate-and-forward strategy is analytically solved using the
replica symmetric ansatz which is exact in the studied system at
the Nishimori's temperature. In particular, the typical level of
improvement in communication performance by relaying messages is
shown in the case of small and large number of relay units.
\end{abstract}


\title{The typical behaviour of relays}
\author{Alamino, R.C., Saad, D.}
\affiliation{Neural Computing Research Group, Aston University, Birmingham, United Kingdom}
\maketitle

\section{Introduction}

Methods of statistical mechanics have recently become
increasingly more important in the study of communication
channels. The development of the replica and cavity methods for
analysing disordered systems~\cite{MPV,Nishimori01} and the
related recent introduction of systematic rigorous
bounds~\cite{Guerra03,Franz01} made new theoretical tools available
for their analysis.

More specifically, the replica method has been applied to a wide
range of problems in information theory, from error correcting
codes~\cite{Sourlas89,ksRev} to multiuser
communication~\cite{Tanaka02}. It facilitates the derivation of
practical and theoretical limits in various communication channels
and provides typical results in cases that are difficult to tackle
via traditional methods of information theory.

The growing use of information networks, both physically connected
and wireless, and the increasing number of services taking place
in the Internet, have made the study of multiuser communication
highly attractive and relevant from a practical point of view, in
addition to being a challenging and exciting field for theoretical
research.

Up to date, there is no generalised theory of multiuser channels
within the framework of information theory and analytical results
are only known for special cases; the main difficulty being that
multiuser networks do not admit the source-channel separation
principle which plays an essential role in the analysis of
communication channels. Nevertheless, multi-user communication
plays an important role in a variety of communication devices
ranging from mobile phones to computers. We strongly believe that
a statistical physics-based analysis may offer answers where the
current information theory methodology fails, especially in the
limit of a large number of users.

With the technological demand and the possibility of a more
thorough study by the methods of statistical mechanics, early
results for multi-user communication are being revisited and
analysed from different and complementary points of view,
resulting in new insights and
developments~\cite{Tanaka02, NakamuraBroad}. One of those
interesting types of channels is the \emph{relay
channel}~\cite{Cover91}. This generic channel is characterised by
an auxiliary user between the transmitter and receiver, which
helps in the transmission of the message. Due to the increase in
the number of multi-user networks, like mobile phones and computer
networks, the transfer of information with the help of a relays
became an attractive option. As these networks are becoming more
distributed, the transmission with the help of arrays of relays
becomes feasible and merits further analytical exploration.

This paper is organised as follows. In
section~\ref{section:System} we define the general relay array and
introduce as particular cases the classical relay channel and the
relay-without-delay. In section~\ref{section:Decoding} we outline
the statistical physics methods used to analyse the problem which
will be based on a replica approach detailed in
section~\ref{section:RS}. Section~\ref{section:Conclusions}
contains our conclusions and final comments.

\section{The Model}
\label{section:System}

\subsection{LDPC Codes}

Low-Density Parity-Check (LDPC) codes~\cite{Gallager62} are
state-of-the-art error-correcting codes with performance that is
second to none, especially within the high code rate regime. In
the notation we will be using here, $N$-dimensional messages
$\mb{s}$ are encoded into $M$-dimensional codewords $\mb{t}$. LDPC
codes are defined by a binary \emph{parity-check matrix} $A=[C_1
\mid C_2]$, concatenating two very sparse matrices known to both
sender and receiver: $C_2$ that is invertible and of
dimensionality $(M-N)\times(M-N)$ and $C_1$ of dimensionality
$(M-N)\times N$. The matrix $A$ can be either random or regular,
characterised by the number of non-zero elements per row ($K$) and
column ($C$). Irregular codes show superior performance to regular
structures~\cite{Richardson01,idosaad1} if constructed carefully.
In order to simplify our treatment, we focus here on regular
constructions; the generalisation to irregular codes is
straightforward~\cite{Vicente02,VSK}.

\emph{Encoding} refers to the linear mapping of a $N$-dimensional
original message $\mb{s} \in \{0,1\}^N$ to a $M$-dimensional
codeword $\mb{t}\in\{0,1\}^M$ ($M>N$)
\begin{equation}
\label{eq:encoding} \mb{t} = G \mb{s} \ \ \mbox{(mod 2)} \ ,
\end{equation}
where all operations are performed in the field $\{0,1\}$ and are
indicated by $\mbox{(mod 2)}$. The generator matrix is
\begin{equation}
\label{eq:generator} G = \prs{\begin{array}{c} I \\  C_{2}^{-1}C_{1}
\end{array}} \ \ \mbox{(mod 2)} \ ,
\end{equation}
where $I$ is the $N\times N$ identity matrix. By construction $A G
= 0 \ \mbox{ (mod 2)}$ and the first $N$ bits of $\mb{t}$
correspond to the original message $\mb{s}$.

\emph{Decoding} is carried out by estimating the most probable
transmitted vector from the received corrupted
codeword~\cite{Vicente02,ksRev}. For mathematical convenience, in
the present work we map the Boolean variable $\mb{t}\in\{0,1\}^M$
into a spin variable $\mb{t}\in\{1,-1\}^M$ by the transformation
$x\rightarrow (-1)^x$.

\subsection{The Relay Array}

\begin{figure}
\includegraphics[width=7cm]{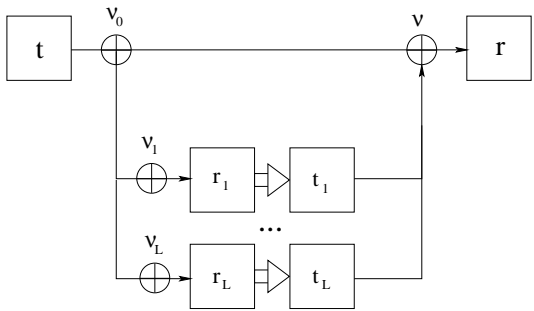}
\caption{\label{figure:RA}
         The $L$-component relay array. The original sender sends a codeword $\mb{t}$
         to the final receiver and to each of the $L$ relays. Each relay receives
         a message $\mb{r}_i$ which is a corrupted version of the original codeword
         subject to the AWGNs $\gb{\nu}_0$ and $\gb{\nu}_i$. It then sends to the final
         receiver the encoded vector $\mb{t}_i$. The final receiver receives the
         original transmitted codeword summed with all $\mb{t}_i$'s and corrupted by
         the AWGN vectors $\gb{\nu}_0$ and $\gb{\nu}$.}
\end{figure}

The \emph{relay array} is a many-units generalisation of the
(single unit) relay channel of~\cite{Cover91}. The LDPC codeword
$\mb{t}$ is transmitted to each one of $L$ relay units through
noise channels corrupted by a global Additive White Gaussian Noise
(AWGN) $\gb{\nu}_0$ and by local independent AWGNs $\gb{\nu}_i$.
Each relay processes the received corrupted message $\mb{r}_i$ and
encodes the acquired information into a vector $\mb{t}_i$ which is
then transmitted to a final receiver. The final receiver receives
an algebraic summation of the relay outputs plus a direct
transmission from the original sender, corrupted also by
$\gb{\nu}_0$, subject to a final AWGN $\gb{\nu}$. The exact form
of the channel is depicted in Fig.~\ref{figure:RA} and the
corresponding equations are
\begin{align}
  \mb{r}   & = a\mb{t}+\sum_{i=1}^L b_i \mb{t}_i+\gb{\nu}+\gb{\nu}_0,\\
  \mb{r}_i & = c_i\mb{t} + \gb{\nu}_i+\gb{\nu}_0.
\end{align}

The variables $a$, $b_i$ and $c_i$ ($i=1,...,L$) are the relative
gains of each transmission and can be random or set to constant
values. The power from the original source to each relay is
$c_i^2$, to the final receiver is $a^2$ and the power from each
relay to the final receiver is $b_i^2$.

When $L=1$, we refer to the channel simply as the \emph{relay
channel}. In the \emph{classical relay channel} (CRC), studied by
Cover and El-Gamal~\cite{Cover79}, the messages sent by the relays
to the final receiver are only allowed to depend on the set of
symbols received by the each of the relays \emph{before} the
current time step, $t_i^\mu=f(t^1,...,t^{\mu-1})$, which
corresponds to the fact that it takes the relay some time to
process the information before relaying it. However, if the time
delay in the direct transmission to the final receiver is much
longer than in the transmission to the relay units, we can allow
the message sent by the relay to depend on the present received
symbol as well such that $t_i^\mu=f(t^1,...,t^\mu)$. This last
case, termed \emph{relay-without-delay} (RWD), created significant
interest recently and was studied by El-Gamal and
Hassanpour~\cite{ElGamal05}. For the case of a relay array where
all communication is carried out through the relays and there is
no direct transmission to the final receiver, the restriction of
the CRC, to consider all but the last received symbol, is
unnecessary.

\section{Statistical Physics of Decoding}
\label{section:Decoding}

We transform the decoding problem of the final receiver into a
statistical physics system by defining a dynamical variable
$\gb{\tau}\equiv(\tau^1,...,\tau^M)$ which represents the
candidate variable vectors at the receiver. Each $\tau^\mu$ plays
a role equivalent to a spin located in the $\mu$-th site of a
lattice with $M$ sites.

The final receiver generates an estimate $\hat{\mb{t}}$ of the
original codeword using the Marginal Posterior Maximiser (MPM)
estimator
\begin{equation}
  \hat{t}^\mu= \sgn\avg{\tau^\mu}{\prob{\gb{\tau}|\mb{r}}},
\end{equation}
which minimises the probability of bit
error~\cite{Iba99,Vicente02}. Other estimators can be used
depending on the error measure considered. For example,
minimisation of \emph{block error} is obtained using the
 Maximum a Posteriori (MAP) estimator.

The posterior probability density is calculated by Bayes' rule as
\begin{equation}
  \label{equation:Inference}
  \prob{\gb{\tau}|\mb{r}}= \frac{\prob{\mb{r}|\gb{\tau}}\prob{\gb{\tau}}}{\prob{\mb{r}}},
\end{equation}
with
\begin{align}
  \prob{\mb{r}} & = \sum_{\mb{t}} \prob{\mb{r}|\mb{t}}\prob{\mb{t}} \nonumber\\
                & = \sum_{\mb{t},\chs{\mb{t}_i},\chs{\mb{r}_i}} \prob{\mb{t}} \prob{\mb{r}|\mb{t},\chs{\mb{t}_i}}
                    \prod_i \prob{\mb{t}_i|\mb{r}_i} \prob{\mb{r}_i|\mb{t}}.
\end{align}

One of the basic quantities of interest is the overlap between the
codeword and the decoded message. Our analysis, focuses on the
typical behaviour of the decoding process and, accordingly, we
take averages over all possible codewords, all received messages
and all allowed encodings, which we consider as quenched disorder
in the system. The overlap between decoded and original message
takes the form
\begin{equation}
   d = \frac1M \sum_{\mu=1}^M \avg{t^\mu \,
   \sgn\avg{\tau^\mu}{\prob{\gb{\tau}|\gb{r}}}}{\iav}.
\end{equation}

This quantity can be derived from the free-energy
\begin{equation}
  f=-\lim_{M\rightarrow \infty} \frac1{\beta M} \avg{\ln Z}{\iav},
\end{equation}
with the partition function
\begin{equation}
  Z=\sum_{\mb{t}} e^{-\beta \mathcal{H}\prs{\mb{t};\mb{r}}},
\end{equation}
and the corresponding Hamiltonian
\begin{equation}
  \mathcal{H}\prs{\mb{t};\mb{r}} = -\ln \prob{\mb{r}|\mb{t}}\prob{\mb{t}}.
\end{equation}

Usually we disregard the normalisation of the distributions within
the Hamiltonian as they merely add constants that shift the zero
energy. In the case of LDPC codes, $\prob{\mb{t}}$ turns out to be
a constraint on the summation variables.

In the above Hamiltonian, the parity-check matrix $A$ defines an
interaction between the $\tau$ variables while $\mb{t}$ and
$\mb{r}$ act as local fields. The inverse temperature $\beta$
is the ratio between the true and the decoder's assumed noise
level. In our numerical calculations, we adopt $\beta=1$, also
known as \emph{Nishimori's temperature}, which means that the
decoder assumes the correct noise level for the channel. It can be
shown that at Nishimori's temperature the system never enters the
glassy phase~\cite{Nishimori01, Montanari01} and the
thermodynamically dominant solution is always Replica Symmetric
(RS); we therefore restrict our analysis to the RS treatment.

One of the important properties and the novelty of the statistical
physics formulation of the problem is that looking at the problem
as a dynamical spin system, we can interpret the results in terms
of phase transitions which can be characterised by the overlap and
the entropy function. Combining this extra information we can have
a better understanding of the way the system changes from a phase
of perfect decoding (which is called the \emph{ferromagnetic
phase}) to a phase where the message is recovered only up to a
certain amount of error (the \emph{paramagnetic phase}).

The most studied strategies used by the relay units are the
Amplify-and-Forward (A\&F) and the Decode-and-Forward (D\&F)
strategies. In A\&F, the relay just retransmits its received
vector, i.e., $\mb{t}_i=\mb{r}_i$. As the replica treatment of
this case turns out to be the same as for the simple Gaussian
channel with a modified power and noise level, the solution is
obtained straightforwardly by applying the results
of~\cite{Vicente02} and will not be studied here.

In the D\&F strategy, the relays decode the message and transmit
their estimate to the final receiver. Full use of LDPC decoding in
the relays is made when each relay decodes the received vector
$\mb{r}_i$ by the MPM estimator using the fact that the codeword
was encoded by an LDPC code. The message transmitted to the final
receiver by each relay would then be
\begin{equation}
  \label{equation:Relay_estimative}
  t_i^\mu = \sgn\avg{\tau_i^\mu}{\prob{\gb{\tau}_i|\mb{r}_i}}.
\end{equation}

In equation~(\ref{equation:Inference}) this is equivalent to
setting
\begin{equation}
  \label{equation:D&F_Probability}
  \prob{\mb{t}_i|\mb{r}_i}=\prod_{\mu=1}^M
  \delta\prs{t_i^\mu-\sgn\avg{\tau_i^\mu}{\prob{\gb{\tau}_i|\mb{r}_i}}}.
\end{equation}

As $t_i^\mu\in\{\pm1\}$, we can rewrite this probability density as
\begin{equation}
  \prob{\mb{t}_i|\mb{r}_i}=\prod_{\mu=1}^M \theta
  \prs{\avg{t_i^\mu\tau_i^\mu}{\prob{\gb{\tau}_i|\mb{r}_i}}},
\end{equation}
where $\theta$ is the Heaviside step function.

The replica treatment of the LDPC D\&F turns out to be extremely
involved due to the introduction of a theta function with an
average over the variables $\gb{\tau}_i$ inside it, which includes
a term dependant on the parity-check matrix. Analytical studies in
order to solve this, rather difficult case, are under way.

In the present work we focus on a simplification of this strategy
also known in the literature as Demodulate-and-Forward. In it
case, the relays do not have the complete information about the
encoding mechanism and therefore assume a uniform prior for the
transmitted codeword. In this case, the posterior distribution of
the bits in the message for the relay is
\begin{equation}
  \prob{\mb{t}_i|\mb{r}_i}=\prod_{\mu=1}^M
  \frac1{1+\exp\col{-2t_i^\mu r_i^\mu/(\sigma_i^2+\sigma_0^2)}},
\end{equation}
and it is straightforward to show that the MPM estimator is given
simply by
\begin{equation}
  t_i^\mu = \sgn\prs{r_i^\mu}.
\end{equation}

The fact that the disorder relative to different codes does not
appear in the estimate of the relays makes the replica
calculations feasible in this case, as follows.

\section{Replica Symmetric Analysis}
\label{section:RS}

As the RS analysis of LDPC coding systems has been introduced and
carried out in a number of publications (e.g. ~\cite{Vicente02})
we will skip the detailed derivation and concentrate on the final
expressions. The derivation follows exactly the same steps as
in~\cite{Alamino07} where quenched averages are first carried out,
followed by the RS assumption which enables the representation of
the order parameters in the form of field distributions. These are
obtained using a set of self-consistent saddle point equations of
the form
\begin{align}
  \label{equation:saddle_point}
  \hpi (\hx) & = \avg{\delta\prs{\hx-\prod_{m=1}^{K-1}x^m}}{\bx},\\
  \pi(x)     & = \avg{\delta \prs{x-
                 \frac{\sum_\tau \tau\,\col{\Psi(\tau,r)}^\beta\prod_{l=1}^{C-1} \prs{1+\tau\hx^l}}
                      {\sum_\tau \col{\Psi(\tau,r)}^\beta\prod_{l=1}^{C-1}\prs{1+\tau\hx^l}}}}{r,\bhx},
                      \nonumber
\end{align}
where
\begin{align}
  \Psi(\tau,r) & \equiv \int \chs{\prod_{i=1}^L dr_i \,\exp\col{-\frac{ \prs{r_i-c_i\tau}^2}{2(\sigma_i^2+\sigma_0^2)}}}
                 \nonumber\\
               & \times \exp\col{-\frac1{2(\sigma^2+\sigma_0^2)}\prs{r-a\tau-\sum_i b_i\,\sgn
               r_i}^2}.
\end{align}
The expression $\avg{t_i}{}$ is the mean of the variable $t_i$ and
$\prob{r} \propto \Psi(1,r)$. The overlap is
\begin{align}
\label{eq:overlap}
  d        &= \avg{\sgn u}{u}, ~\mbox{with}\\
  \prob{u} &= \avg{\delta \prs{u-
              \frac{\sum_\tau \tau\,\col{\Psi(\tau,r)}^\beta\prod_{l=1}^C \prs{1+\tau\hx^l}}
              {\sum_\tau \col{\Psi(\tau,r)}^\beta\prod_{l=1}^C\prs{1+\tau\hx^l}}}}{r,\bhx},
\end{align}
the free energy is given by
\begin{align}
  \beta f & =  \frac{C}K \ln 2 +C\avg{\ln(1+x\hx)}{x,\hx}\nonumber\\
          &   -\frac{C}K\avg{\ln\prs{1+\prod_{m=1}^K x^m}}{\bx}\nonumber\\
          &   -\avg{\ln\chs{\sum_\tau \col{\Psi(\tau,r)}^\beta\prod_{l=1}^C\prs{1+\tau\hx^l}}}{\bhx,r},
\end{align}
and the internal energy, the derivative with respect to $\beta$ of the above equation, is
\begin{align}
  u & =  -\avg{\frac{\sum_\tau \Psi^\beta (\ln \Psi) \prod_{l=1}^C\prs{1+\tau\hx^l}}
                     {\sum_\tau \Psi^\beta\prod_{l=1}^C\prs{1+\tau\hx^l}}}{\bhx,r}.
\end{align}

For any number $L$ of relays, the results can be obtained by a
numerical solution of the equations~(\ref{equation:saddle_point}).
Note the summation over the internal variables, i.e., the messages
received and sent by the relays. This comes from the Bayesian
formulation of the problem where the final receiver has access
just to $\mb{r}$ and, therefore, must integrate over the unknown
variables.

We also note that the above equations are fairly general. Using
the appropriate ``potential'' $\Psi$ we can recover all previous
results for single user channels and apply them to more general
channels when the intermediate processing of the message does not
involve the previous knowledge of the parity-check matrices.

The ferromagnetic state, which corresponds to perfect decoding, is
given by the following solution to the saddle point equations
(\ref{equation:saddle_point})
\begin{equation}
  \hpi(\hx)=\delta(\hx-1), \qquad \pi(x)=\delta(x-1).
\end{equation}

Substitution of these distributions in equation~(\ref{eq:overlap})
gives $d=1$. By substituting the ferromagnetic solution into the
formulas for the free and internal energies, we obtain (for
Nishimori's temperature)
\begin{equation}
  u=f=-\avg{\ln \Psi(1,r)}{r},
\end{equation}
meaning that the entropy of this phase is zero.

The Hamiltonian of the relay array is \emph{gauge invariant} with
respect to the gauge transformation
\begin{align}
  r^\mu &\rightarrow \gamma^\mu r^\mu, \nonumber\\
  t^\mu &\rightarrow \gamma^\mu t^\mu,
\end{align}
where the vector $\gb{\gamma}$ obeys the parity-check constraints.
We can verify that the transition probabilities
$\prob{r^\mu|t^\mu}$ are also invariant under this gauge
transformation. Note that if a channel is symmetric (for a
definition see~\cite{Tanaka03}), it is automatically gauge
invariant under the above transformation. For gauge invariant
channels the internal energy is
\begin{align}
  U &= \avg{\mathcal{H}\prs{\gb{\tau};\mb{r}}}{\gb{\tau},\mb{r},\mb{t}} \nonumber\\
    &= \sum_{\gb{\tau},\mb{t}}\int d\mb{r}\, \prob{\gb{\tau}|\mb{r},\beta}\prob{\mb{r}|\mb{t}}\prob{\mb{t}}
       \mathcal{H}\prs{\gb{\tau};\mb{r}},
\end{align}
where
\begin{equation}
  \prob{\gb{\tau}|\mb{r},\beta} \propto e^{-\beta \mathcal{H}},
\end{equation}
is the thermal Gibbs probability at inverse temperature $\beta$
which obeys
$\prob{\gb{\tau}|\mb{r},\beta=1}=\prob{\gb{\tau}|\mb{r}}$. Since
under such a gauge transformation the Hamiltonian remains
invariant, we have
$\mathcal{H}\prs{\mb{t};\mb{r}}=\mathcal{H}\prs{\mb{1};\mb{t}\mb{r}}$,
where $\mb{t}\mb{r}\equiv (t^1 r^1,..., t^M r^M)$ and $\mb{1}$ is
an $M$-dimensional vector with all entries equal to 1. Therefore,
one can write
\begin{equation}
  U  = \sum_{\gb{\tau},\mb{t}}\int d\mb{r}\,
       \frac{\prob{\mb{r}|\gb{\tau},\beta}\prob{\gb{\tau}|\beta}}{\prob{\mb{r}|\beta}}
       \prob{\mb{r}|\mb{t}}\prob{\mb{t}}
       \mathcal{H}\prs{\mb{1};\gb{\tau}\mb{r}}.
\end{equation}

Gauging the variables $\gb{\tau}\mb{r}\rightarrow\mb{r}$,
reorganising the terms and taking $\beta=1$, we finally get
\begin{equation}
  U = \int d\mb{r}\, \prob{\mb{r}|\mb{1}} \mathcal{H}\prs{\mb{1};\mb{r}}.
\end{equation}

The meaning of this is that, for a gauge invariant channel of the
type described above (which includes the symmetric channels), the
internal energy is independent of the configuration. In special
cases, as can be found in~\cite{Nishimori01}, the gauge symmetry
allows an analytical expression to be found. The same method can
be used to prove that the probability distribution for the
magnetisation is equal to the probability distribution for the
two-point correlations in Nishimori's temperature, which indicates
the absence of a spin glass phase and no replica symmetry
breaking.

\subsection{The Relay Channel}

\begin{figure}
\includegraphics[width=8.6cm]{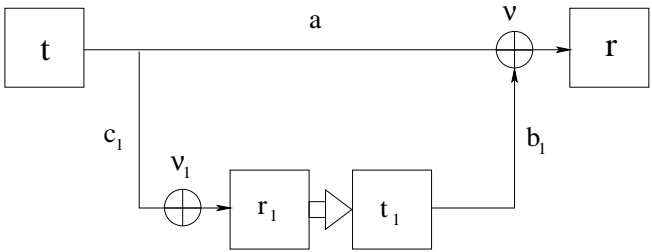}
\caption{\label{figure:RWD_setup} Schematic drawing of the relay-without-delay (RWD) setup to be analysed.}
\end{figure}

In order to compare our results with those of~\cite{ElGamal05}, we
analyse the RWD for the setup sketched in Fig.~\ref{figure:RWD_setup}
with $\sigma_1^2=\eta \sigma^2$, $a=b_1=1$ and
$c_1=(1+\sigma^2)^{-1/2}$. The potential is then given by
\begin{align}
  \Psi(\tau,r) & = e^{-\prs{r-\tau-1}^2/2\sigma^2}\,\erfc\prs{-\frac{\tau}{\sqrt{2\eta\sigma^2}}}\nonumber\\
           & + e^{-\prs{r-\tau+1}^2/2\sigma^2}\,\erfc\prs{+\frac{\tau}{\sqrt{2\eta\sigma^2}}},
\end{align}
where $\erfc(x)$ is the complementary error function
\begin{equation}
  \erfc(x)= \frac{2}{\sqrt{\pi}} \int_x^{\infty} e^{-y^2}\,dy.
\end{equation}

For these values of noise and gains, the capacity of this channel
as derived in~\cite{ElGamal05} is
\begin{equation}
  \mathcal{C}=\frac12 \log_2 \prs{1+\frac{1+c_1^2}{\sigma^2}}.
\end{equation}

\begin{figure}
\includegraphics[width=8.6cm]{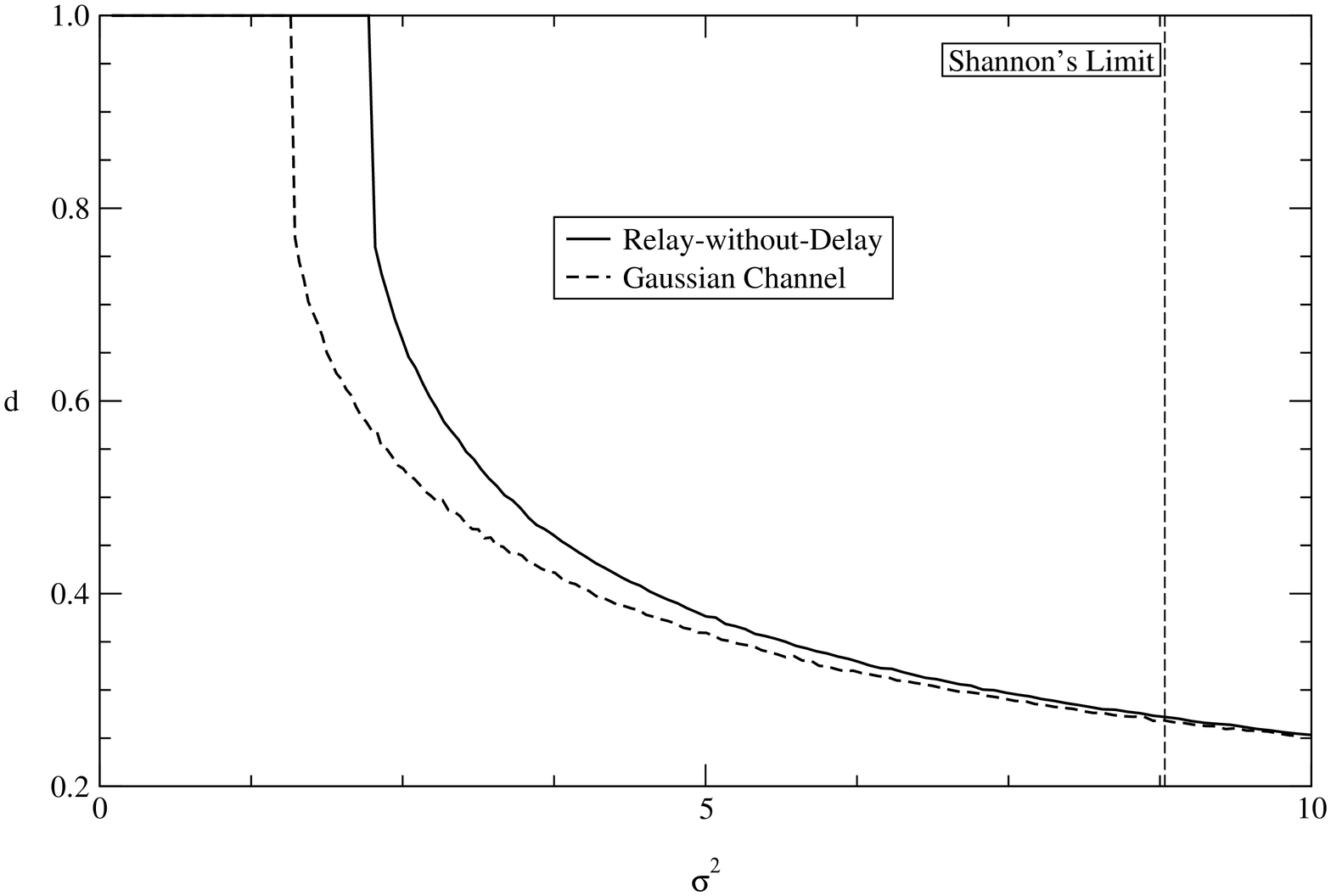}
\caption{\label{figure:RWD_overlap} The overlap of retrieved and
        original codewords for the RWD channel with
         $\sigma_1^2=0.1\sigma^2$ is given by the continuous curve. The dashed curve
         shows the same for the simple Gaussian channel. The vertical line indicate
         Shannon's limit for the RWD as calculated by El-Gamal and
         Hassanpour~\cite{ElGamal05}.}
\end{figure}

The numerical results for the overlap between the retrieved and
the original codewords, obtained by solving recursively
equations~(\ref{equation:saddle_point}), are given in
Fig.~\ref{figure:RWD_overlap} for $K=4$, $C=3$, $\beta=1$ and
$\eta=0.1$. Shannon's limit is indicated by the vertical dashed
line and corresponds to a noise level $\sigma^2\approx 8.79$. The
dashed curve shows the overlap for a simple Gaussian channel with
noise level $\sigma^2$ and the continuous one shows the overlap
for the RWD. The improvement in the practical limit for error-free
communication, indicated by the highest noise level for which
$d=1$ is clear. However, the distance between the dynamical
transition threshold $\sigma_d^2 \approx 2.22$ and Shannon's limit
for the channel is greater than in the case of the simple Gaussian
channel (for numerical results for the Gaussian channel
see~\cite{Tanaka03}). Numerical calculations point to the expected
result that decreasing the noise level from the source to the
relay brings $\sigma_d^2$ closer to Shannon's limit. However, one
must remember that the relay strategy examined does not use the
full potential of the relay and the additional information
embedded in the LDPC codes. We expect that a LDPC decoding
\emph{in the relay} will improve the communication performance and
currently focus on the analysis of this scenario.

We can also see in the plot that, as the noise level increases,
the channel becomes closer to the Gaussian channel. This is just a
consequence of the fact that, for high noise level, the additional
information provided by the relay becomes negligible as both users
decode the message poorly.

\begin{figure}
\includegraphics[width=8.3cm]{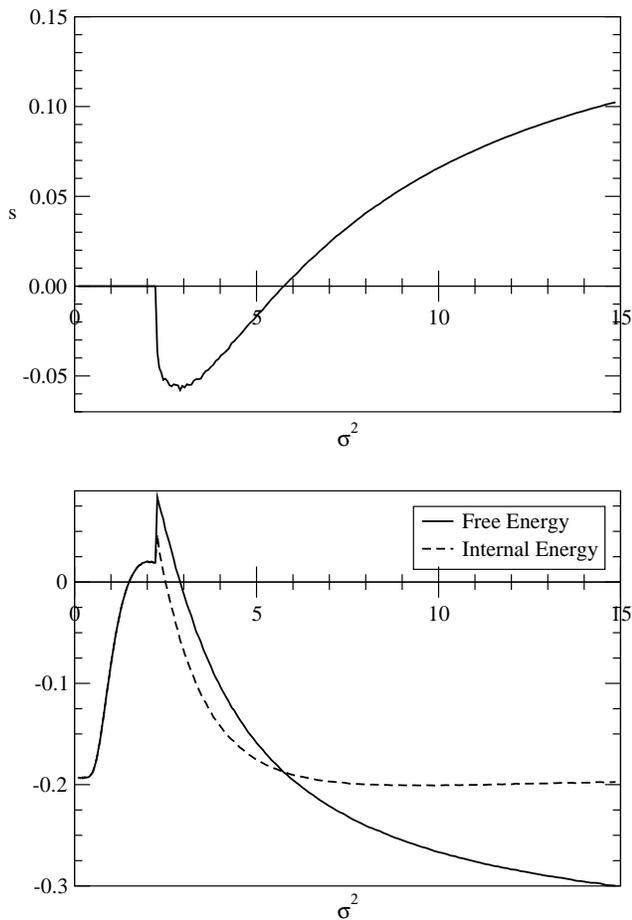}
\caption{\label{figure:RWD_entropy} Entropy and energies for the
RWD channel at Nishimori's temperature. The upper graph shows the
entropy which is given by the difference between the internal
energy and the free energy depicted in the bottom graph.}
\end{figure}

Figure~\ref{figure:RWD_entropy} shows the entropy and the free and
internal energies for the same values as in
Fig.~\ref{figure:RWD_overlap}. At the dynamical transition point,
where practical perfect decoding becomes unfeasible, the entropy
becomes negative, indicating the emergence of subdominant
metastable states that can be further explored using the replica
symmetry breaking ansatz. Between this point and the
thermodynamical transition point, where the entropy is positive
again, the dominant state is still ferromagnetic but the
population dynamics algorithm used to solve the saddle point
equations becomes trapped in a local minimum with free-energy
higher than the ferromagnetic one. Due to the equality between the
internal energy and the ferromagnetic free-energy, the point where
the entropy becomes positive again is also the point where both
energies cross in the bottom graph.

\begin{figure}
\includegraphics[width=8.3cm]{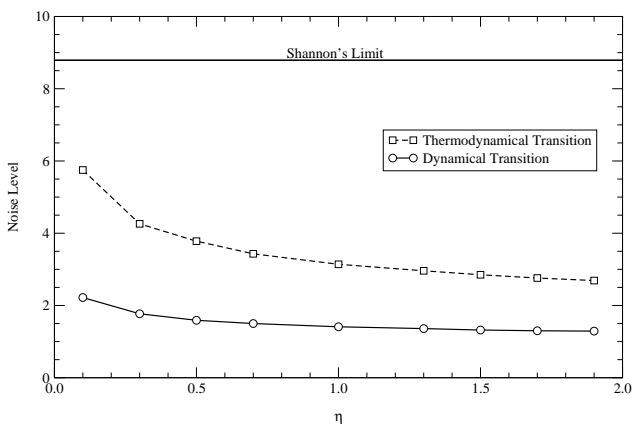}
\caption{\label{figure:RWD_eta} The continuous/dashed line shows
the dynamical/thermodynamical transition noise levels of the RWD
against $\eta$, the ratio between relay and final receiver noise
levels. The upper horizontal line corresponds to Shannon's limit
$\sigma^2\approx 8.79$.}
\end{figure}

In Fig.~\ref{figure:RWD_eta} we plot the dynamical and
thermodynamical transition noise levels against $\eta$, the ratio
between the relay and final receiver noise levels. We see that the
dynamical and thermodynamical transition points decrease with
$\eta$ but became closer to each other, stabilising at asymptotic
values that match those of the simple AWGN channel values as the
relay contribution becomes meaningless.

\begin{figure}
\includegraphics[width=8.6cm]{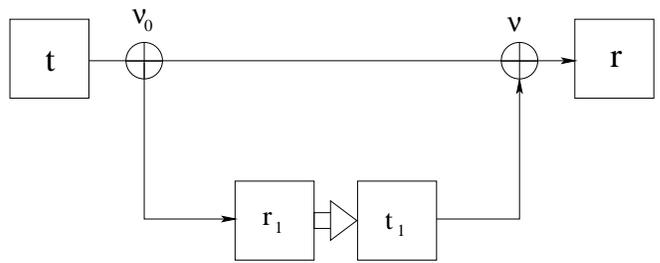}
\caption{\label{figure:CRC_setup} Schematic drawing of the
classical relay channel (CRC) setup. The relative gains  are all
equal to 1 and not shown in the picture.}
\end{figure}

Although the capacity for the RWD is known only in special cases,
its upper bound can be higher than in the case of the CRC. In
order to verify it for the LDPC-based framework we analyse in this
paper, we now use a setup equivalent to the one studied
in~\cite{Cover79} and shown in Fig.~\ref{figure:CRC_setup} where
$a=c_1=b_1=1$ and $\sigma_0^2\equiv \lambda\sigma^2$. The capacity
of the CRC in this case is
\begin{equation}
  \label{equation:CRC_capacity}
  \mathcal{C} =
  \left\{
  \begin{array}{cc}
    \frac12 \log_2\prs{1+\frac1{\lambda \sigma^2}}, & \lambda\geq1, \\
     & \\
    \frac12 \log_2\prs{1+\frac4{(1+\lambda)^2 \sigma^2}}, & \lambda<1.
  \end{array}
  \right.
\end{equation}

\begin{figure}
\includegraphics[width=8.6cm]{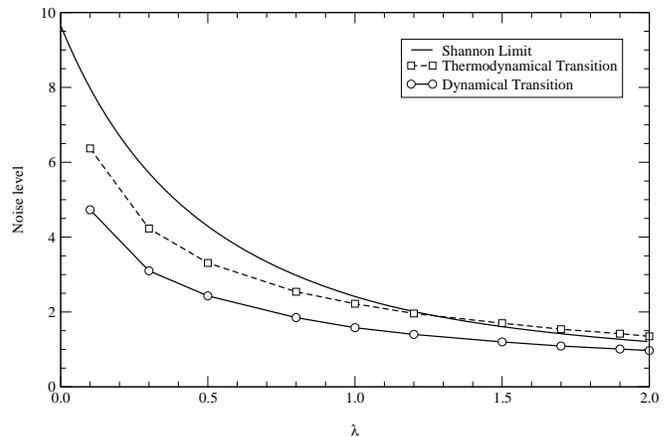}
\caption{\label{figure:CRC_lambda} The continuous/dashed line
shows the dynamical/thermodynamical transition noise levels of the
RWD against $\lambda$ in the setup of Fig.~\ref{figure:CRC_setup}.
The continuous line without marked symbols is Shannon's limit for
a CRC with the same noise levels and transmission powers.}
\end{figure}

In Fig.~\ref{figure:CRC_lambda} we compare the dynamical and
thermodynamical threshold noise levels of a RWD with Shannon's
limit for the CRC, both with the setup described above, for
different values of $\lambda$, the ratio between the noise levels
applied at the transmission and reception points.

We can see that, although the practical decoding line (dynamical
transition) falls below Shannon's limit for all calculated values, the
thermodynamical transition goes above it \emph{for
the CRC case} at higher values of $\lambda$.
Figure~\ref{figure:CRC_lambda} shows that the capacity of the RWD
is indeed higher than the CRC for the case studied and quantifies
the gain in allowing the message sent by the relay to depend on
the current transmitted symbol (which is excluded in the CRC).
Although allowing this instantaneous dependence would at first
sight seem just a small modification which is insignificant in the
infinite block length limit, it indeed gives relevant extra
information which facilitates more efficient retrieval at the
final receiver. The insight gained is that for the RWD and large
$\lambda$, the relay transmission $t_1^\mu$ is correlated with the
original codeword $t^\mu$, which is not the case in the CRC
allows; this allows for an improvement in the information
extraction at the receiver.

\subsection{Large Relay Array}
\label{sTLRA}

Now, we will use the central limit theorem to obtain the result for
large $L$ in the relay array setup given by Fig.~\ref{figure:RA}.
As the relay messages are correlated and to guarantee that the
quantities have the same order, we introduce a $1/L$ scaling in
the summation over relay messages. The potential for this model
becomes
\begin{align}
   &\Psi(\tau,r)  \equiv \int \chs{\prod_{i=1}^L dr_i \,\exp\col{-\frac{ \prs{r_i-\tau}^2}{2(\sigma_i^2+\sigma_0^2)}}}
                  \nonumber\\
                & \times \exp\col{-\frac1{2(\sigma^2+\sigma_0^2)}
                  \prs{r-\tau-\frac1L\sum_i \sgn r_i}^2},
\end{align}
where we assumed, for simplicity, $a=b_i=c_i=1$.

For $L\gg 1$, the central limit theorem amounts to a modification
in the distribution of the variable $r$ given by
\begin{equation}
  \prob{r} = \int \col{\prod_{i=1}^L dr_i\,\prob{r_i}} F\prs{\frac1L\sum_{i=1}^L \sgn r_i}
           = \avg{F(\omega)}{\omega},
\end{equation}
where
\begin{align}
  F(\omega)  &= \frac1{\sqrt{2\pi\sigma^2}}\exp\col{-\frac{\prs{r-1-\omega}^2}
                {2(\sigma^2+\sigma_0^2)}},\\
  \prob{r_i} &= \frac1{\sqrt{2\pi\sigma_i^2}}\,\exp\col{-\frac{\prs{r_i-1}^2}{2(\sigma_i^2+\sigma_0^2)}},\\
\end{align}
with
\begin{equation}
  \prob{\omega}=\norm{\frac1L\sum_{i=1}^L \avg{\sgn r_i}{r_i}}{\frac1{L^2}\sum_{i=1}^L \prs{1-\avg{\sgn r_i}{r_i}}^2}.
\end{equation}

For simplicity, we consider the case where the noise level is the
same for all relays $\sigma_i^2=\sigma_1^2$ and define
\begin{equation}
  \sigma_r^2\equiv \sigma^2_1+\sigma_0^2, \qquad \sigma_f^2\equiv \sigma^2+\sigma_0^2.
\end{equation}

Then the corresponding distribution for $\omega$
\begin{equation}
  \prob{\omega} = \norm{\erf\prs{1/\sqrt{2}\sigma_r}}{\frac1L\,\erfc^2\prs{1/\sqrt{2}\sigma_r}}.
\end{equation}

Consequently the contribution for the final noise level coming
from the relay transmission decreases as $L^{-1}$. In the limit
$L\rightarrow\infty$, this distribution becomes a delta function
centred at the error function and therefore
\begin{equation}
  \prob{r} = \norm{1+\erf\prs{1/\sqrt{2}\sigma_r}}{\sigma_f^2}.
\end{equation}

Accordingly, the potential becomes
\begin{equation}
   \Psi(\tau,r) = \exp\chs{-\frac1{2\sigma_f^2}\col{r-\tau-\erf\prs{\tau/\sqrt{2}\sigma_r}}^2}.
\end{equation}

\begin{figure}
\includegraphics[width=8.6cm]{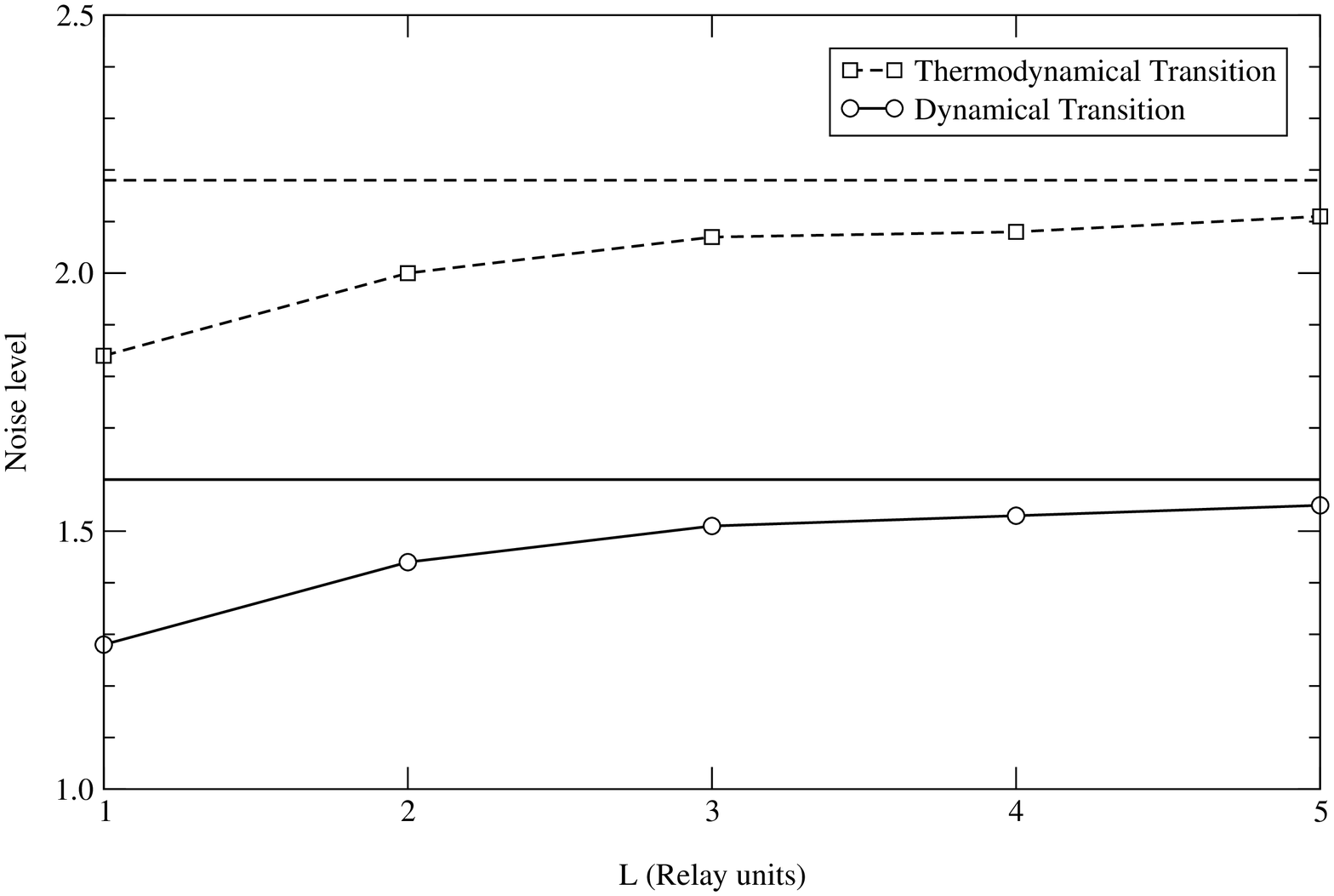}
\caption{\label{figure:RWD_L} Dynamical and thermodynamical
transition points for many relays. The exact formula is used to
calculate the points $L=1,2,3,4,5$. The horizontal lines represent
the large $L$ limit.}
\end{figure}

Figure \ref{figure:RWD_L} compares the dynamical and
thermodynamical transition points for $L=1,2,3,4,5$ calculated by
the exact formula and the result obtained by the approximation for
large $L$. Again, we consider the case of $K=4$, $C=3$ and
$\beta=1$.

It is clear from Fig.~\ref{figure:RWD_L} that already at $L=5$,
both dynamical and thermodynamical transition points approach the
large $L$ limit solution, thus making this approximation
attractive already for low $L$ values.

\section{Conclusions}
\label{section:Conclusions}

In this work we analysed the behaviour of relay arrays using
methods of statistical mechanics. These networks are of growing
significance due to the increase of multi-user, mobile and
distributed communication systems.

We found an analytical solution for the relay-without-delay (RWD)
channel given by the RS ansatz, which due to the gauge symmetry of
the channel, is exact at Nishimori's temperature that correspond
to a choice of the correct prior within the Bayesian framework. We
showed the level of improvement with respect to a simple Gaussian
channel without relaying which, even for the naive relay strategy
of Demodulate-and-Forward analysed here, is significant.

We compared the RWD dynamical and thermodynamical transition
points for different noise ratios between the relay and the direct
channel; and found that although these points become far from
Shannon's limit, the difference between the dynamical and the
thermodynamical transition decreases. The relevance of the relay
is clearly decreasing as its noise level increases as the level of
additional information it conveys diminishes.

We also were able to compare the RWD case to the classical relay
channel (CRC) for different noise ratios between the relay and the
direct channel. We found that  the capacity of the RWD is higher
than the CRC for a high relay noise, showing the significance of
the extra information conveyed by the relay on the current
transmitted symbol, which is absent in the CRC framework.

The performance of a large array of relays was analysed and
compared against results obtained for a small number of units. The
result obtained are consistent and indicate that this useful
approximation provides accurate results already for a small number
of units. For a large array, we also found that the increase in
noise tolerance levels off.

We have demonstrated the usefulness of methods adopted from
statistical physics for analysing multi-user communication
systems. While we have concentrated on limited scenarios of relay
channels, we believe that these methods hold a promising
alternative to the information theory methodology which, in
general, has not been successful in dealing with multi-user
communication systems. The study of different relay channels and
other multi-user communication networks is underway.

\section*{Acknowledgements}
Support from EVERGROW, IP No. 1935 in FP6 of the EU is gratefully acknowledged.

\bibliographystyle{prsty}
\bibliography{inftheory,statphysecc}

\end{document}